\documentstyle[12pt]{article}

\def\binom#1#2{{#1 \choose #2}}

\def\beq{\begin{equation}}
\def\eeq{\end{equation}}
\def\beqa{\begin{eqnarray}}
\def\eeqa{\end{eqnarray}}
\def\bega{\begin{array}}
\def\enda{\end{array}}

\let\theta=\theta

\def\ssty{\scriptstyle}
\def\sssty{\scriptscriptstyle}

\def\frac#1#2{{#1\over#2}}
\def\frc#1#2{\relax\ifmmode{\textstyle{#1\over#2}} 
                    \else$#1\over#2$\fi}           
\def\frcc#1#2{\relax\ifmmode{\sssty{#1\over#2}} 
                    \else$#1\over#2$\fi}           

\def\rlx{\relax\leavevmode}                  
\def\inbar{\vrule height1.5ex width.4pt depth0pt}
\def\sinbar{\vrule height1ex width.35pt depth0pt}
\def\ssinbar{\vrule height.7ex width.3pt depth0pt}
\font\cmss=cmssi12
\font\cmsss=cmssi12 at 7pt
\def\ZZ{\rlx\leavevmode
             \ifmmode\mathchoice
                    {\hbox{\cmss Z\kern-.4em Z}}
                    {\hbox{\cmss Z\kern-.4em Z}}
                    {\lower.9pt\hbox{\cmsss Z\kern-.36em Z}}
                    {\lower1.2pt\hbox{\cmsss Z\kern-.36em Z}}
               \else{\cmss Z\kern-.4em Z}\fi}
\def\Ik{\rlx{\rm I\kern-.18em k}}  
\def\IC{\rlx\leavevmode
             \ifmmode\mathchoice
                    {\hbox{\kern.33em\inbar\kern-.3em{\rm C}}}
                    {\hbox{\kern.33em\inbar\kern-.3em{\rm C}}}
                    {\hbox{\kern.28em\sinbar\kern-.25em{\rm C}}}
                    {\hbox{\kern.25em\ssinbar\kern-.22em{\rm C}}}
             \else{\hbox{\kern.3em\inbar\kern-.3em{\rm C}}}\fi}
\def\IP{\rlx{\rm I\kern-.18em P}}
\def\ZZ{\rlx{\rm Z\kern-.25em Z}}
\def\IR{\rlx{\rm I\kern-.18em R}}
\def\Ident{\rlx{\rm 1\kern-2.7pt l}}
\def\ID{\relax{\rm I\kern-.18em D}}
\def\IF{\relax{\rm I\kern-.18em F}}
\def\IH{\relax{\rm I\kern-.18em H}}
\def\II{\relax{\rm I\kern-.17em I}}
\def\IN{\relax{\rm I\kern-.18em N}}
\def\IQ{\relax\,\hbox{$\inbar\kern-.3em{\rm Q}$}}

\def\ie{\hbox{\it i.e.}}        

\def\gsim{\relax\leavevmode
     \ifmmode\mathchoice			
	{\raise1pt\hbox{$>$} \kern-0.65em \lower3.5pt\hbox{$\ssty \sim$}}
	{\raise1pt\hbox{$>$} \kern-0.65em \lower3.5pt\hbox{$\ssty \sim$}}
	{\raise0.75pt\hbox{$\ssty >$} \kern-0.56em
		\lower3.5pt\hbox{$\sssty \sim$}}
	{\raise0.85pt\hbox{$\sssty >$} \kern-0.50em
		\lower2.7pt\hbox{$\sssty \sim$}}
     \else				
	{$\raise1pt\hbox{$>$} \kern-0.65em
		\lower3.5pt\hbox{$\ssty \sim$}$}\fi}

\def\lsim{\relax\leavevmode
     \ifmmode\mathchoice			
	{\raise1pt\hbox{$<$} \kern-0.65em \lower3.5pt\hbox{$\ssty \sim$}}
	{\raise1pt\hbox{$<$} \kern-0.65em \lower3.5pt\hbox{$\ssty \sim$}}
	{\raise0.75pt\hbox{$\ssty <$} \kern-0.56em
		\lower3.5pt\hbox{$\sssty \sim$}}
	{\raise0.85pt\hbox{$\sssty <$} \kern-0.50em
		\lower2.7pt\hbox{$\sssty \sim$}}
     \else				
	{$\raise1pt\hbox{$<$} \kern-0.65em
		\lower3.5pt\hbox{$\ssty \sim$}$}\fi}

\begin{document}

\author{Luis Masperi, Ariel M\'egevand \\
Centro At\'omico Bariloche and Instituto Balseiro\\
Comisi\'on Nacional de Energ\'{\i}a At\'omica and\\
Universidad de Cuyo, 8400 S. C. de Bariloche, Argentina}
\title{Stability of modified electroweak strings }
\date{}
\maketitle

\begin{abstract}
We discuss the stability of an electroweak string with axions in its core,
which give to the configuration a quasi-topological property, and compare it
with other modifications  using instantons in the thin-wall approximation.
\end{abstract}

\vspace{1cm}


\pagebreak
\section{Introduction}

The existence of electroweak strings might be of relevance both for the
explanation of cosmological problems and future accelerator results \cite
{V-V}. However, when they are built exclusively with Higgs and gauge fields
of the standard model we obtain classically unstable configurations, or at
most quantum metastable ones with significant life-time only for too small
Higgs boson mass \cite{JPV} and too large Weinberg angle \cite{KO}.

To give more stability to the electroweak string, an additional global
abelian symmetry may be considered. When this symmetry is unbroken, its
Noether charge for particles lighter inside than outside the core may
produce stable strings as occurs for non-topological solitons \cite{V-VW}.
Alternatively, if the global symmetry is broken, there is a topological
reason for the stability of a configuration where the electroweak component
is added to an axionic string \cite{DS}, though QCD effects may cause its
decay. We propose a closed string, that partially combines both effects,
with axions in its core which are lighter than outside and a
quasitopological basis for stability due to the phase nature of the axion
field that allows a variation of a multiple of $2\pi $ along it. Absolute
stability is not obtained due both to the very small decay probability of
axions and to the finiteness of the transverse region of this coherent
configuration inside the core.

In section 2 we describe the reasons of instability  of electroweak strings
and attempts to stabilize them. Section 3 is devoted  to the proposed string
with axions in its core showing that the configurations might be macroscopic
and with large life-time. Brief conclusions are included in section 4.

\section{Electroweak strings and attempts of stabilization}

We remind that the stability of vortices relies on the non-triviality of $
\Pi _1(M)$, where $M$ is the vacuum manifold, and eventually on the
conservation of the magnetic flux through the core in the gauge theories. In
the case of a broken global $U(1)$ symmetry, the stability of the so called
axionic string is assured by $\Pi _1(U(1))=Z$. For the Nielsen-Olesen vortex
\cite{NO} corresponding to the breaking of a local $U(1)$ symmetry,  if this
$U(1)$ was the remnant of a larger broken symmetry $G$ such that $\Pi
_2(G/U(1))\neq I$  a quantum decay of the configuration to a broad one may
occur\cite{PV}.

The limit $g=0$ of the electroweak string is the semilocal one \cite{H-AKPV}
which corresponds to the breaking $SU(2)_{gl}\times U(1)_{loc}\rightarrow
U(1)_{gl}$ with a complex Higgs field ${\binom{\varphi _1}{\varphi _2}}$
where $\varphi _2$ is electrically neutral, a potential

\begin{equation}
\label{e1}V=\frac \lambda 4\left( \left| \varphi _1\right| ^2+\left| \varphi
_2\right| ^2-v^2\right) ^2
\end{equation}
and a neutral gauge field $Z$ with coupling $g^{\prime }$. Though in this
case $\Pi _1(M)=I$, the conservation of $Z$-magnetic flux may produce
stability of the vortex. In fact being the configuration such that for large
$r$ in $D=2$

\begin{equation}
\label{e2}\varphi \rightarrow {\binom 0v}e^{i\theta },\quad Z_\theta
\rightarrow \frac 1r
\end{equation}
and for small  $r\quad \varphi \rightarrow 0,\quad Z\rightarrow 0$, if we
increase $\varphi _1$ up to $v$ inside the core and then enlarge its size $
R\rightarrow \infty $, the remaining energy is the additional scale
invariant gradient contribution $\sim v^2$. The vortex will be stable if its
potential and magnetic energies $\sim \sqrt{\lambda }v^2/g^{\prime }=\frac{
M_H}{M_Z}v^2$ are smaller than the previous gradient contribution, being
this possible if the Higgs mass is lighter than the vector boson
one\cite{PV}.  If on the contrary $M_H>M_Z$ the vortex becomes metastable
tunneling to a broader configuration with the same energy. Finally, for a
sufficiently large $M_H\sim 2M_Z$ the radius of the latter configuration
becomes equal to the original vortex size, there is no barrier to surmount
and the defect is unstable. A channel for decay of the $D=3$ string is its
breaking by nucleation of a global monopole pair.
Again for large $M_H\sim 2M_Z$ the string will become
unstable when the energy balances for a pair separation of the order of the
defect size so that there will be no barrier to surmount.

In the true electroweak case $g\neq 0$, $\tan \theta =g^{\prime }\!/g\,$,
the $B_Z$-flux is not conserved and may transform into the electromagnetic
$B_A$-flux \cite{N}. Therefore the vortex can never be stable because one
may change the configuration eliminating the phase of $\varphi _2$ in
Eq.(\ref{e2} ) by a gauge transformation, then turning $B_Z$-flux into
$B_A$-flux which is decoupled from $\varphi _2$ thus avoiding a divergent
energy contribution when $\varphi _2$ is subsequently increased to $v$
inside the core. No gradient contribution is introduced in this way and
enlarging $R\rightarrow \infty $ to eliminate the magnetic energy, one
obtains a vacuum configuration of zero energy without having surmounted any
infinite barrier.  For $D=3$ one decay channel is given by the nucleation of
magnetic monopoles with the probability $P$ given by the bounce action $B$
\cite{PV}

\begin{equation}
\label{e3}P\sim e^{-B},\quad B\sim \frac{E_{mon}^2}{\mu _s}
\end{equation}
Analogously to the semilocal case, the string energy per unit length is
\linebreak $\mu _s\sim \sqrt{\lambda }\sin \theta v^2/g^{\prime }=\frac{M_H}{
M_Z}v^2$ and the monopole energy for the realistic case $g>g^{\prime }$ can
be estimated by the sum of the contribution of a global monopole for a small
range $<1/M_Z$ and that of the magnetic energy emerging from it i.e.
\begin{equation}
\label{e4}E_{mon}\simeq \frac{v^2}{M_Z}+\frac{\sin {}^2\theta }{g^2}M_Z=
\frac{M_Z}{g^2}
\end{equation}
The energy balances for a separation of monopoles which is larger than their
size, condition for metastability, for large $\theta $ and small $\frac{M_H}{
M_Z}$ excluding therefore the realistic region. The same conclusion is
reached for $D=2 $ analizing the monopole pair bounce configuration
responsible for the decay of the $Z$-string into a $B_A$-flux configuration.
It is important to remark that the instability of $Z$-strings is
additionally increased by its possible decay into a $W^{+}W^{-}$ condensate
\cite{MT}. This is due to the vacuum instability caused by the anomalous
coupling of the spin of $W$ with $A$ and $Z$ magnetic fields \cite{AO}.
In particular $W$s are produced with a zero energy expense for a strong $Z$
magnetic field \begin{equation}
\label{e5}B_Z \gsim \frac{M_W^2}{g\cos
\theta }
\end{equation}
which, for the flux inside a $Z$-string, is satisfied for $M_H \gsim M_Z$.

To obtain a larger stability one may think of including an additional global
symmetry $\tilde U(1)$. If this symmetry is unbroken there is no topological
reason for stability since one still has $\Pi _1(M)=I$ but the conservation
of its Noether charge may stabilize the configuration as occurs in
non-topological solitons. Thus an example is to introduce a complex scalar
field $\chi $ with the additional lagrangian \cite{V-VW}
\begin{equation}
\label{e6}{\cal L}_\chi =\left| \partial _\mu \chi \right| ^2-\lambda
_2\left| \chi \right| ^4-\lambda _3\left( \varphi ^{\dagger }\varphi
-m_0^2\right) \left| \chi \right| ^2,\quad v^2>m_0^2
\end{equation}
with parameters such that the mass of the particle $\chi $ is smaller inside
the vortex core, where there is an expectation value $\chi_0$, than its mass
$m_\chi =\sqrt{v^2-m_0^2}$ outside. For $D=2$ we may pass from the vortex
configuration in whose core $\varphi \sim 0,\;\left| \chi \right| \simeq
\chi _0$ to the vacuum $\varphi _2=v,\;\chi =0$ everywhere by the same steps
as in the electroweak case with the only difference that the vortex energy
decreases from $\frac{M_H}{M_Z}v^2$ to $\frac{\sqrt{ M_H^2-\delta
/v^2}}{M_Z}v^2$, where $\delta $ corresponds to the depth of the local
minimum of the potential caused by Eq.(6). The replacement of the Higgs mass
by $ \sqrt{M_H^2-\delta /v^2}$ in all the above considerations tends to
increase the life-time of the metastable configurations and to allow them
for more realistic values of $M_H$. For $D=3$ it would be possible to obtain
absolute stability if we consider $z,t$ dependent configurations inside the
core corresponding to a Noether charge $Q$ per unit length. If the particles
$ \chi $ are so heavy outside the core that $Qm_\chi >\tilde \mu $, where
$\tilde \mu $ is the energy per unit string length including the $Q$
particles inside it, the string cannot decay. Otherwise for the monopole
nucleation channel the bounce action of Eq.($\ref{e3}$) is modified
through the replacement of $\mu $ by $\tilde \mu -Qm_\chi $ with the
possibility of increasing the life-time.

It is interesting to consider
the realistic case in which the fermions that acquire their mass through
coupling with the Higgs field may be trapped in the string core. It has been
shown that in this way the electroweak string becomes superconducting \cite
{MOQ} with zero-mass fermions running along one direction of the string or
the other according to their being an up or a down component of the
elecroweak doublet \cite{W}. But the Dirac sea of these quarks tends to
destabilize the string \cite{Na} because a small perturbation of the upper
Higgs component $ \delta \varphi _1$ produces, via the Yukawa coupling $h$
to the fermion, mixed states for each momentum $p$ with an energy shift
\begin{equation}
\label{e7}\Delta \epsilon =-\left( \sqrt{p^2+\left| h\delta \varphi
_1\right| ^2}-p\right)
\end{equation}
It is therefore necessary that positive energy fermions with a proper
density per unit string are present to overcome this shift and may have a
chance to stabilize the configuration.

One may think to increase the stability to the elecroweak string introducing
topological reasons for it. A possibility is to make the hypothetical
axion field play a role due to its phase nature. A model is built adding the
electroweak components to the axionic string \cite{DS} which corresponds to
a global $\tilde U(1)$ symmetry spontaneously broken at a higher scale
$f_{PQ}$ so that $$ \Pi _1\left( SU(2)\times U(1)\times \tilde
U(1)/U(1)_{em}\right) =\Pi _1\left( \tilde U(1)\right) \neq I.  $$
To have a
coupling between the phase of the scalar field $\psi $ which breaks this
Peccei-Quinn symmetry with the Higgs sector one needs two complex Higgs
doublets $\varphi _i$ and an effective potential \begin{equation} \label{e8}
\begin{array}{r} V(\varphi _i,\psi )=\sum_{i=1}^2 \frac{\lambda _i}4\left(
\varphi _i^{\dagger }\varphi _i-v_i^2\right) ^2+\frac \lambda 2\left(
\varphi _1^{\dagger }\varphi _1\right) \left( \varphi _2^{\dagger }\varphi
_2\right) + \\  \\ +\frac{\lambda ^{\prime }}2\left( \varphi _1^{\dagger
}\varphi _2\right) \left( \varphi _2^{\dagger }\varphi _1\right)
+f_{PQ}v_1v_2-\frac 12\left[ \left( \varphi _1^{\dagger }\varphi _2\right)
\psi +h.c.\right] \end{array} \end{equation}
Eq.($\ref{e8}$) forces the difference of phases of the Higgs doublets to be
equal to that of $\psi$ outside the string core i.e. $\theta _1-\theta
_2=\theta _\psi$ with \begin{equation} \label{e9}\varphi _i={\binom
0{v_i}}e^{i\theta _i},\quad \psi =f_{PQ}e^{i\theta _\psi }
\end{equation}
A modification emerges from QCD corrections which give a small
mass $m_a$ to the axion breaking intrinsically the symmetry $\tilde
U(1)\rightarrow Z(N)$, where in the invisible axion model $N$ is the number
of quark flavours.  Therefore the string becomes attached to $N$ axionic
walls which have too much energy for cosmology \cite{DS2}. If
the attached wall is a single one it becomes surrounded by the string
turning into a metastable configuration \cite{PV}. When the electroweak
components are included into the axionic string, the $Z$-flux is conserved
due to Eq.(\ref{e9}) adding stability to the configuration because its
electroweak part returns to the situation of the semilocal case.

\section{Electroweak string with axions in its core}

We propose a different mechanism for the stabilization of the electroweak
string, inspired in a previously discussed non topological configuration
\cite{FGMM}, which consists in including axions in the core combining the
effects of being lighter there and of corresponding to a phase-like field.
In the case analyzed above one thinks that the skeleton is an axionic string
formed at the Peccei-Quinn scale $f_{PQ}\sim 10^{12}\,GeV$, which attracts
electroweak components at the $SU(2)\times U(1)$ breaking. We consider
instead that the electroweak strings formed at the latter scale trap axions.
While outside the core the real scalar axion field  must be
around zero in correspondence to the vacuum, it takes an average value $a_0$
inside the core where the electroweak symmetry is unbroken. This follows
from the absence of the contribution $\arg \det {\cal M}$ of the mass matrix
\cite{R} so that it must not be compensated by the axion field.

To explain this feature we remind that the vacuum energy $E(\theta)$ of QCD
in a phase where the quarks have mass $m$ depends on the $\bar
\theta$-vacuum parameter as \cite{vw}
\begin{equation}
\label{z1}
\begin{array}{ll}
\exp -\left( E(\bar \theta )VT\right) = & \int
{\cal D}G_\mu \det (\rlap/ D+m) \\  & \cdot \exp -\int d^4x\left( \frac
12G_{\mu \nu }G_{\mu \nu }-i\bar \theta G_{\mu \nu }\tilde G_{\mu \nu
}\right)
\end{array}
\end{equation}
with $G_{\mu \nu}$ gluon fields.

Since $\det (\rlap/ D+m)=m^{N_0}\sum\limits_r\left( \lambda _r^2+m^2\right)
>0$ the factor $e^{i\bar \theta \nu}$ can only decrease the integral so that
$E(\bar \theta )>E(0)$. In case that the axion field is included, all what
said above applies to $\bar \theta -a^{\prime}/f_{PQ}$ which is defined as
$-a/f_{PQ}$. To be precise, $\bar \theta$ contains not only the QCD
parameter  $\theta$ but also $\arg \det {\cal M}$ coming from the EW
breaking.

If we are in the symmetric EW phase and $\det \rlap/ D$ is still positive
$a^{\prime}$ must compensate only $\theta$ and the redefined $a$ will be of
order $f_{PQ}$. If the axion mass in this phase is vanishing in
correspondence to zero modes of $\rlap/ D$ there will not be a preferred
value  of $a$ but the average of the phase $a/f_{PQ}$ will be of order $1$.

What said above may be simulated in the simplest way by an effective
potential
\begin{equation}
\label{e10}V(\varphi ,a)=\frac \lambda 4\left( \varphi ^{\dagger }\varphi
-v^2\right) ^2+\left[ m_a^2f_{PQ}^2+\kappa \left( \varphi ^{\dagger }\varphi
-v^2\right) \right] \left( 1-\cos \frac a{f_{PQ}}\right)
\end{equation}
with $\kappa v^2>m_a^2f_{PQ}^2$.
In the stable
phase $\left| \varphi \right| =v$ Eq.(\ref{e10}) gives the normal axion
potential with mass $m_a$ which forces the field to its minimum $a=0$.
Outside the string we expect this phase  and
inside it, where $\varphi
\sim 0$, the axion mass will be
\begin{equation}
\label{e11}m_a^{^{\prime }2}=\kappa \frac{v^2}{f_{PQ}^2}-m_a^2
\end{equation}
which may be adjusted to the desired extremely small value by the additional
parameter $\kappa$.

We will now show that both decay channels of monopole nucleation and $W$
condensate may be considerably suppressed by this mechanism.

For $D=3$, if we make $a$ vary along the axis of the string particles are
excited and being $m_a^{\prime }\ll m_a$, we obtain an effect equivalent to
that of the Noether charge because of the very long life-time of axions.
But more important than that, due to the fact that $a/f_{PQ}$ is a phase, if
we take a closed string the variation along it must be $ \Delta a=n2\pi
f_{PQ}$ with integer $n$, and there is a quasi-topological reason for an
increase of the stability as it occurs in the superconducting strings
\cite{W}. This follows from the fact that when the string is cut an
additional energy corresponding to a jump of $a$ in the wall width $\epsilon
$ must be supplied.  The decay of the string must be accompanied by an
alignment of the phase $a$ of the field $\psi$ inside the defect with its
value $a=0$ outside in order to minimize the gradient energy.
This alignment may be done smoothly in all the points along the string axis
except for a small region where an abrupt change of phase will appear which
will require to cut the string (see Fig.1) with a gradient energy
of order \begin{equation} \label{e12} d\sim \left(\frac{\pi
f_{PQ}}{\epsilon}\right) ^2 \epsilon \pi R^2 \end{equation} that is not
infinite because of the finiteness of the transverse radius $R$ .

It is important to note that the pure EW string has an energy per unit
length which comes from minimizing essentially the contribution of the
difference of vacua and that of the flux $\Phi$, i.e.
\begin{equation} \label{z2}\mu \simeq \Delta V\pi R^2+\frac{\Phi ^2}{2\pi
R^2} \end{equation} which leads to a field $\vec B_Z$ that is independent of
$R$ and exceeds the critical value Eq.(\ref{e5}) for $M_H>M_Z$ producing the
instability of the configuration.

Now, the addition of axions inside the core  gives a large surface
contribution which together with the flux term determines the energy per
unit length
\begin{equation}
\label{z3}\tilde \mu \simeq \frac{\left( \Delta a\right) ^2}\epsilon 2\pi
R+\frac{ \Phi ^2}{2\pi R^2} \end{equation}
Minimizing Eq.(\ref{z3}) $R^3\sim \epsilon \Phi ^2/\left( \Delta a\right)
^2$, taking the wall width
$\epsilon \sim 1/\sqrt{m_aM_H}\sim 10^6GeV^{-1}$ and being $\Delta a \sim
f_{PQ}$, it is necessary that $\Phi \gsim 10^{18}$ to ensure $R \gsim
\epsilon$. Even though this flux is a large number, if the radius is not too
small $\vec B_Z$ may go below the critical value Eq.(\ref{e5}). In fact with
$R \sim 10^4\epsilon $, $B_Z$ turns out to be $10^{24}G$. Therefore the
string radius must be larger than $10^{-4}cm$ to avoid the instability due
to $W$ condensation.

Regarding the energy balance for the monopole pair
nucleation which may break the string and including its cut contribution
\begin{equation}
\label{e13}
2E_{mon}+d= \tilde \mu L
\end{equation}
the bounce action is now
\begin{equation}
\label{e14}
B\sim
\frac{\left( 2E_{mon}+d\right) ^2}{\tilde \mu }
\end{equation}
If we adopt the above parameters and taking the energy of the monopole
as the magnetic contribution due to $\Phi$ outside a core of the size $R$,
$E_{mon}\sim 10^{38}GeV$ which is of the same order of $d$ coming from
Eq.(\ref{e12}), whereas $\tilde \mu$ Eq.(\ref{z3}) is of order
$10^{28}GeV^2$.  Therefore $B \sim 10^{44}$ to be compared with the
bounce action for a normal EW string $B\lsim 1/g^4\sim 10^2$ for $M_H \gsim
M_Z$, \ie, much more stable.

\section{Conclusions}

We have seen that at variance from other modifications a closed EW string
with the inclusion of axion field in its core which increases continuously
along it may exhibit a very long life-time, even though it is not
completely stable. The high value of the energy of the string is exaggerated
by the estimation of a common wall width for the Higgs and axion fields
which gives a very large surface term for the latter. Therefore for
quantitative applications a more accurate description of the configuration
should be performed. It is clear that one should also study the mechanism
for the production of such heavy and macroscopical objects which
might play a role in cosmological aspects as the baryogenesis during the
electroweak phase transition if this was weakly of first order or of second
order \cite{BD} \cite{FMM}. They might also influence astrophysical
observations as a rotation of the polarization plane of the radiation from
distant radiosources due to the parity-violating coupling with
electromagnetic fields \cite{MS}. This effect caused by coherent axionic
configurations may be more important than that due to the previously
considered background of quasi-Goldstone boson fields \cite{HS}.

\vspace{2cm}

\subsubsection*{Acknowledgements:}

We are deeply indebted to Sandra Savaglio for her collaboration in the early
stage of this work and to F. Klinkhamer for an illuminating
correspondence.

This research was partially supported by CONICET Grant No. 3965/92.


\section*{Figure caption}

\begin{description}
\item[1a]  Closed string with axionic content. Arrows indicate axion phase
changing in $2\pi $ inside the string whereas outside it is fixed to $0$.

\item[1b]  Cut of the string required to align the axion phase to its vacuum
value.
\end{description}

\end{document}